\documentclass[twocolumn,numbook]{svjour3}         
\smartqed  
\usepackage{graphicx}
\usepackage{amsmath} 
\usepackage{longtable}
\usepackage{multicol}
\usepackage{multirow}
\usepackage{booktabs}
\usepackage[numbers]{natbib}
\newcommand{\ignore}[1]{}
\begin{document}

\title{Regulatory networks and connected components of the neutral space 
}
\subtitle{A look at functional islands}


\author{Gunnar Boldhaus \and Konstantin Klemm}


\institute{	G. Boldhaus \and K. Klemm \at
            Bioinformatics Group \\
						Dept. of Computer Science \\
						University of Leipzig \\
						H\"{a}rtelstr. 16-18 \\
						D-04107 Leipzig \\
						Germany \\
            Tel.: +49-341-9716672\\
            Fax: +49-341-9716679\\
            \email{gunnar@bioinf.uni-leipzig.de}\\
            \email{klemm@bioinf.uni-leipzig.de}           
}

\date{Received: date / Accepted: date}

\maketitle

\begin{abstract}
The functioning of a living cell is largely determined by the
structure of its regulatory network, comprising non-linear interactions between regulatory genes. An important
factor for the stability and evolvability of such regulatory
systems is neutrality --- typically a large number of alternative
network structures give rise to the necessary dynamics. 
Here we study the discretized regulatory dynamics of the yeast cell cycle [Li et al., PNAS, 2004] and the 
set of networks capable of reproducing it, which we call \textit{functional}. Among these, the empirical yeast 
wildtype network is close to optimal with respect to sparse wiring. Under point mutations, which establish or 
delete single interactions, the neutral space of functional networks is \textit{fragmented} into $\approx 4.7 \times 10^8$ components. 
One of the smaller ones contains the wildtype network. On average, functional networks reachable from the wildtype by mutations are sparser, 
have higher noise resilience and fewer fixed point attractors as compared with networks outside of this wildtype component.

\end{abstract}

\section{Introduction} \label{intro}
Neutrality \cite{Kimura:83} is crucial for robustness and evolvability \cite{WagnerBook2005}
of biological systems. It describes the fact that the mapping from genotypes to phenotypes
is not invertible. A given phenotype can be encoded by more than one genotype. As Wagner
\cite{WagnerBook2005} writes, {\em ``most problems the living have solved have an astronomical
number of equivalent solutions, which can be thought of as existing in a vast
neutral space''}.

Computational studies of biopolymers revealed the existence of neutrality in the relation
between sequence and spatial structure. RNA molecules and proteins are
generated as a chain (sequence) of nucleic bases and amino acids respectively.
The number of sequences folding into one and the same functionally relevant
spatial structure is found to be large. It is growing exponentially with the size of
the molecule \cite{Schuster:94a,Babajide1997}. Together with an adjacency
given by single mutations, the phenotypically equivalent genotypes
form the neutral network (or {\em neutral graph}). The properties of this
graph, in particular its {\em connectivity}, determine the robustness of the
given genotype under mutations and its evolvability towards new phenotypes.

Going from single molecules to the level of the whole organism, the phenotype
is not given by the set of its molecule structures alone: The dynamics that
arises as the result of activating and suppressing {\em interactions}
between molecules is crucial. This set of interactions is captured as
a regulatory network \cite{Davidson2005} and gives rise to
a temporal sequence of chemical concentration vectors that are responsible for the 
division of a single cell or the development of an embryo. 
Again, the mapping from genotypes (interaction networks) to
phenotypes (temporal sequences) is not injective, i.e. several network
structures are able to produce the regulatory dynamics of a given phenotype
\cite{Ciliberti2007a,Ciliberti2007b}.

Here we apply the neutral graph concept to a dynamical model \cite{Li2004} of
cell cycle regulation in the organism  of the yeast species 
\textit{Saccharomyces cerevisiae} (budding yeast). In section~\ref{sec:model} we
introduce the model dynamics and the wiring of the wildtype network. The
ensemble of functional networks that yield dynamics equivalent to the wildtype
is analyzed in section \ref{sec:functional}, finding the neutral graph to be
disconnected. In section \ref{sec:wildtype} we focus on statistical properties of the subset of networks that
are reachable from the wildtype. After a remark
(section~\ref{sec:computational}) on the computation of network statistics,
section~\ref{sec:discussion} offers a discussion and open questions.

\section{Cell cycle network and Boolean dynamics} \label{sec:model}

During the process of cell division, a eukaryotic cell grows and divides into
two daughter cells. A cell cycle consists of four distinct and separate phases
named $G_1$, $S$, $G_2$ and $M$. In the $G_1$ (''growth'') phase, the cell
commits itself for cell division under certain conditions. In particular, a
necessary cell size must have been reached. A copy of the genetic information is
produced in the $S$ (''synthesis'') phase. The $G_2$ (''gap'') phase precedes the
actual cell division in the $M$ phase (''mitosis'').

\begin{figure*}[ctb]
\centering
\includegraphics[width=0.99\textwidth]{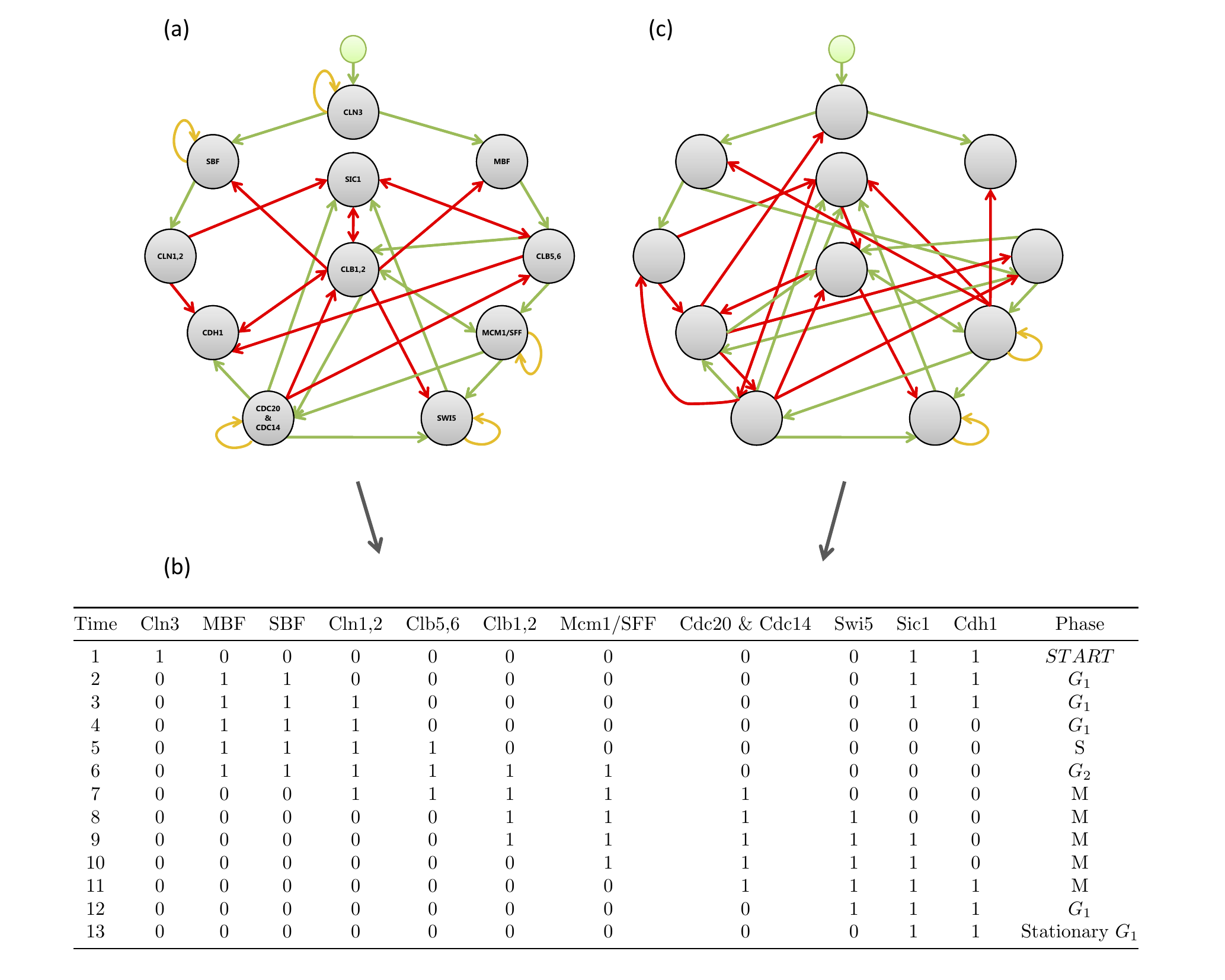}
\caption{\label{fig:wiring}
(a) The Cell Cycle Network of the yeast wildtype has 11 nodes 
connected with activating (green) and inhibiting (red) interactions.
Self-suppression is indicated by yellow loops.
(b) A sequence of 13 states defines a cell cycle, as produced by the network in (a).
(c) A different network (mutant) performs the same sequence of states.
As the wildtype, this mutant has 34 interactions. However, 19 entries in the
interaction matrix differ from the wildtype.}
\end{figure*}

Here we are interested in the network of molecules (cyclins,
inhibitors and degraders of cyclins and transcription factors) regulating
this process. We consider the regulatory network of the
monocellular eukaryotic organism \textit{Saccharomyces cerevisiae} (budding
yeast). Its genome comprises 13 million base pairs and $6275$ genes, of which
approximately 800 are involved in the cell cycle dynamics \cite{Spellman1998}.
The dynamics is controlled by a core of $11$ key regulators
with $34$ directed interactions \cite{Li2004}, shown in Figure 
\ref{fig:wiring}(a), which we denote as the {\em wildtype} network.
Interactions are captured by a matrix $A$. If node
$j$ has an activating effect on node $i$, the corresponding matrix element
is $a_{ij}=+1$, while inhibition is coded as $a_{ij}=-1$. In case of no direct
influence from $j$ to $i$, we have $a_{ij}=0$. Li et al.\ \cite{Li2004} model 
the regulatory dynamics with a Boolean approach \cite{Kauffman1969,Drossel2008} where
each node $i$ takes state values $S_i (t)\in \{0,1\}$ when being inactive / active 
at time $t$. In the time-discrete dynamics, nodes are updated synchronously, 
based on their weighted input sum $h_i(t) = \sum_{j} a_{ij} S_{j}(t)$. The
state at the next time step is obtained by applying the threshold update rule
\begin{equation}
S_{i}(t+1)=\left\{
\begin{aligned}
1 \text{,}& & h_i(t) > 0  \\
0 \text{,}& & h_i(t) < 0 \\
S_{i}(t)  \text{,}& & h_i(t) = 0 
\end{aligned}\right. ~.
\label{eq:dynamics}
\end{equation}
From an initial condition $S(1)$, representing the real starting state of the
cell cycle, the dynamics produces the sequence of state vectors
$S(1),S(2),\dots,S(13)$, shown in Figure \ref{fig:wiring}(b). The state $S(13)=G_{1}$
is a fixed point of the dynamics. The system remains in this state until node
Cln3 is externally activated. In the real system the external activation
indicates that the cell size is sufficient for another division.

\section{Functional networks and the neutral graph} \label{sec:functional}

Broadening our treatment of regulatory networks, we consider the set of
all networks with interaction matrices over $11$ nodes with entries
$a_{ij} \in \{-1,0,+1\}$. We call a network {\em functional} if it produces
the state transitions of the cell cycle sequence in Figure \ref{fig:wiring}(b).
Thus, the wildtype network is functional. However, there are further 
functional networks. Out of the set of all
$3^{2^{11}} \approx 5.4 \times 10^{57}$ networks,
approximately $5.11 \times 10^{34}$ are functional \cite{Lau2007}.
Figure \ref{fig:wiring}(c) shows an example of a functional network
different from the wildtype.

\begin{figure}
\includegraphics[width=0.50\textwidth]{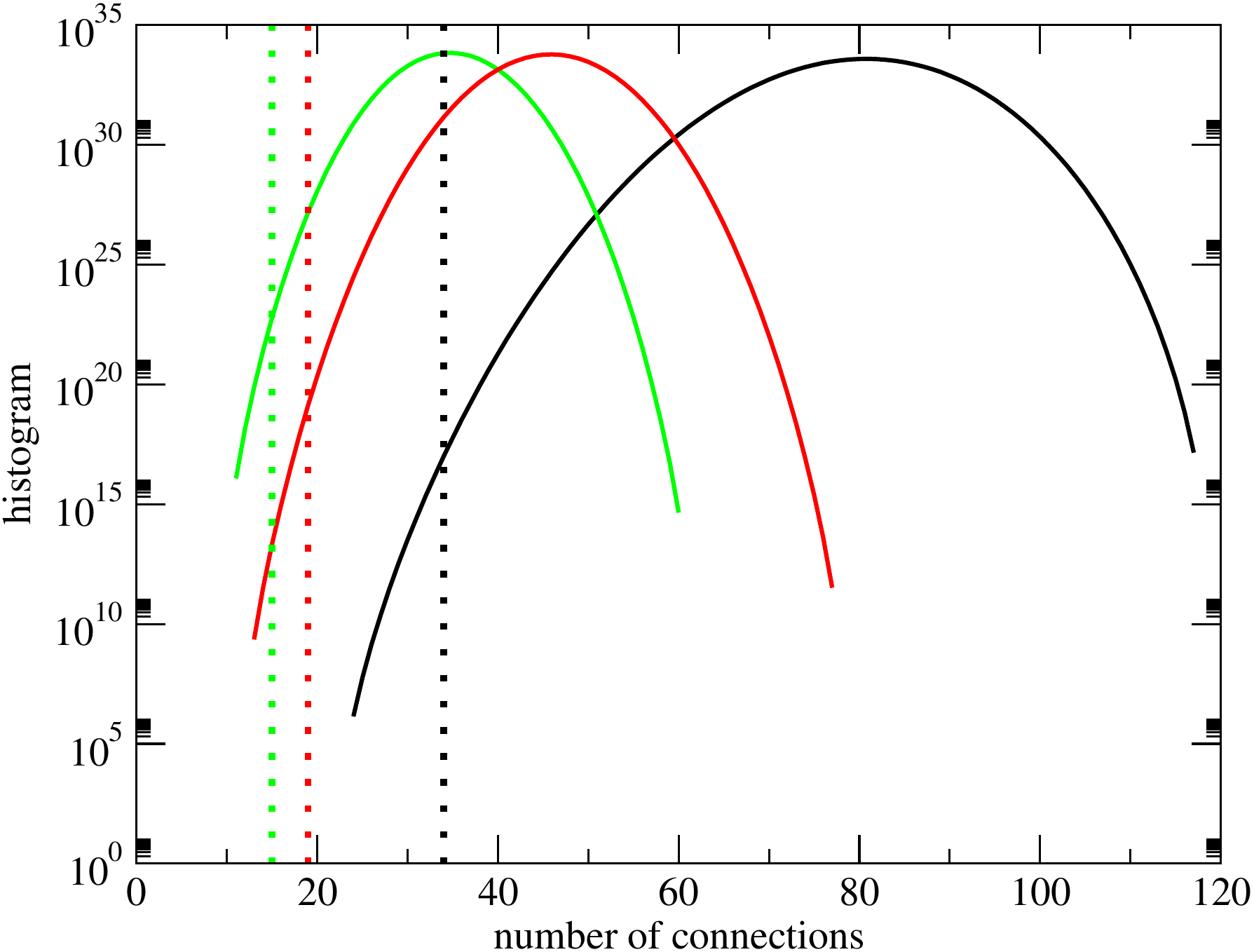}
\caption{\label{fig:hist}       
Histograms of the number of interactions over functional networks.
Positive (green curve), negative (red curve),
curve) and total connections (black curve) of almost all functional networks
exceed the corresponding counts in the wildtype network (vertical dashed lines).}
\end{figure}

Figure \ref{fig:hist} shows the statistics for the number of interactions
(arcs) present in functional networks. The wildtype network is sparse in comparison
with the average functional network. However, there are functional networks
that are even sparser than the wildtype. These findings analogously hold when
activating and inhibiting interactions are counted separately. Interestingly,
functional networks have generally more suppressing than activating
interactions, as is the case for the wildtype.

A structure to reflect mutations on the set of functional networks is the {\em neutral
graph}. Its nodes are the functional networks. Functional networks $A$ and
$B$ are adjacent (connected by an edge) in the neutral graph if $A$ is turned
into $B$ by a single {\em mutation}. According to our definition a mutation is a replacement of
one entry in the interaction matrix. The Hamming distance between two
networks is the number of entries in which their interaction matrices differ.
In order to avoid confusion with the networks of interaction we employ the term {\em neutral graph} as a
synonym for the more commonly used {\em neutral network}.
\begin{figure}
\includegraphics[width=0.50\textwidth]{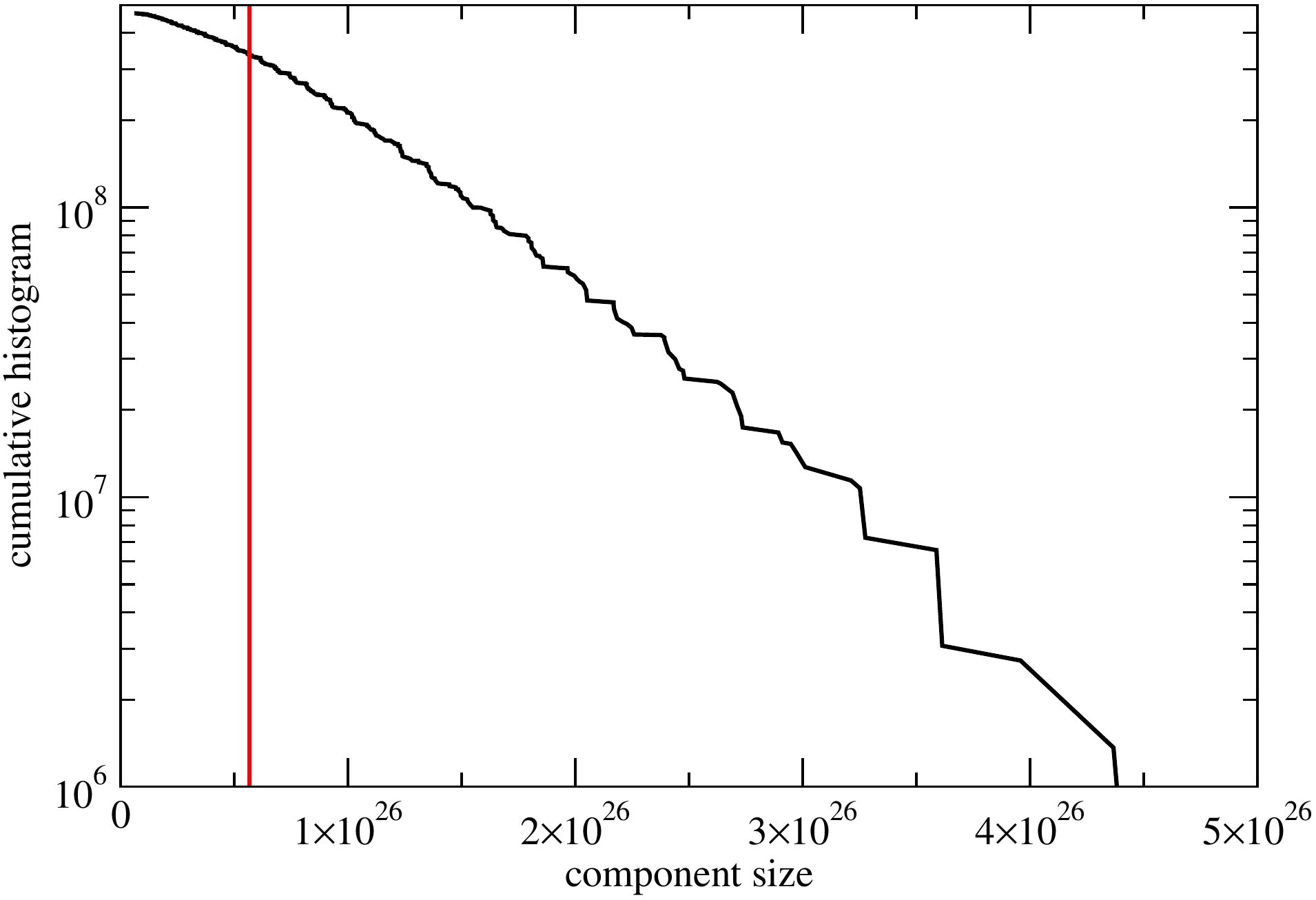}
\caption{Cumulative size distribution of connected components of the neutral
graph (falling curve). The component containing the wildtype has size $5.66 \times 10^{25}$
(vertical line).}
\label{fig:cumhistcomp}       
\end{figure}
An important property of a neutral graph is its connectedness. 
A mutational walk from network $A$ to network $B$ is a sequence of single point mutations that turns $A$ into $B$ without passing through non-functional networks. The neutral graph is connected if such a mutational walk exists for each pair of functional networks.
We find that the neutral graph considered here is {\em disconnected}. One cannot
pass from all functional networks to all others by sequences of mutations that
preserve functionality. In fact, mutual reachability between functional networks
is rare. The neutral graph falls into $\approx 4.7 \times 10^8$ connected
components with sizes distributed between $\approx 6.1 \times 10^{24}$ and
$\approx 4.4 \times 10^{26}$, as shown in Figure \ref{fig:hist}. The component
of the wildtype comprises around $5.66 \times 10^{25}$ functional networks.

\section{The wildtype component} \label{sec:wildtype}
\begin{figure*}
\centerline{\includegraphics[width=0.90\textwidth]{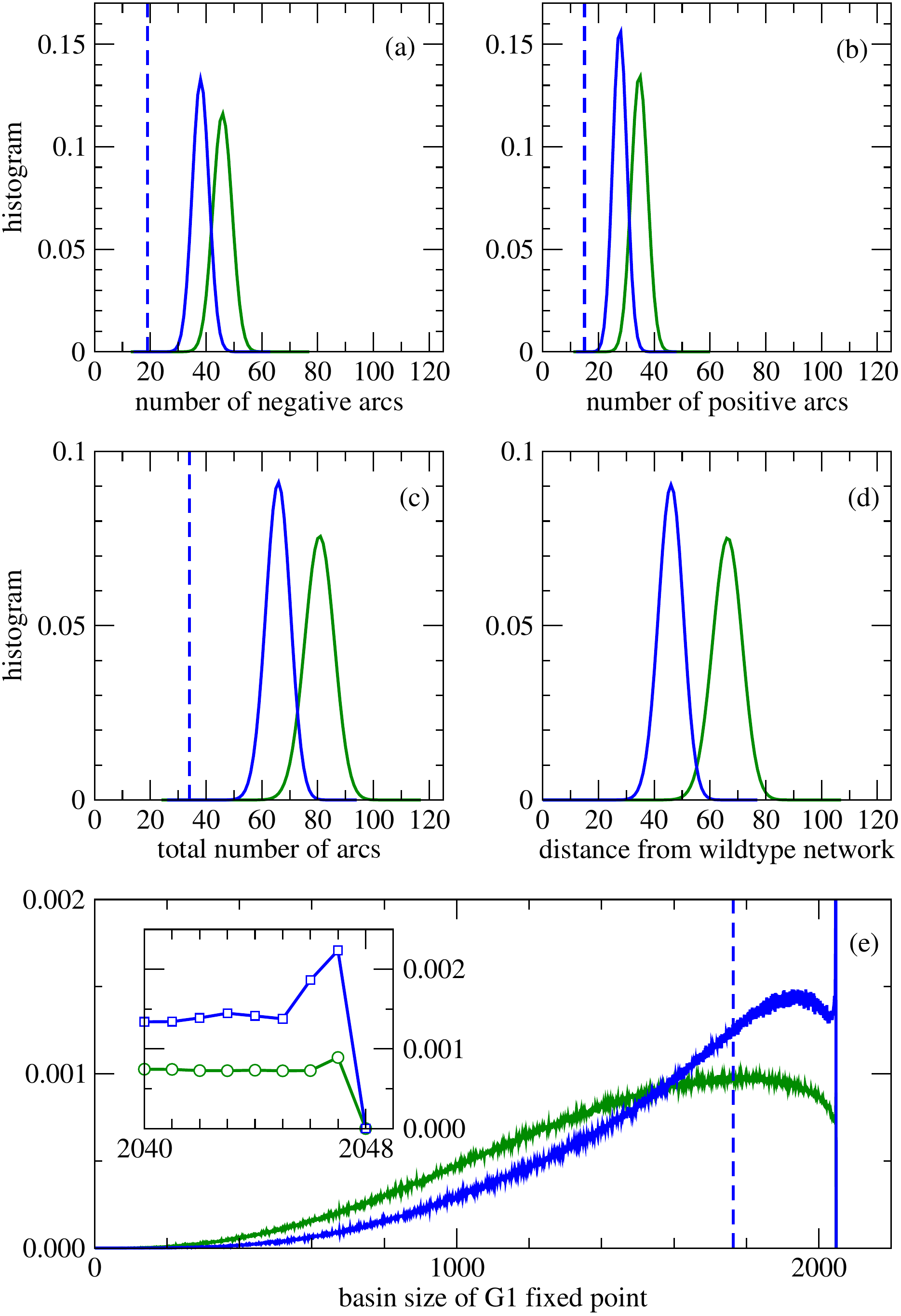}}
\caption{\label{fig:allhist_comp_0}       
Comparison of statistics between all functional networks (green curves) and functional networks 
in the wildtype component (blue curves) of the neutral graph. (a) histogram of negative interactions, 
(b) histogram of positive interactions and (c) histogram of total number of interactions in functional networks. 
(d) histogram of Hamming distances (minimal number of mutations) from the wildtype. 
(e) Distribution of basin sizes of the $G_1$ fixed point. The inset shows a zoom into the histogram for very large basin sizes. 
Histograms in panels (a)-(d) are exact. Histograms in (e) were obtained by uniform sampling of $10^6$ functional networks each 
from the whole neutral graph and from its wildtype component, respectively.}
\end{figure*}

In this section we extend the analysis of the neutral graph. We focus on a comparison between
functional networks in the wildtype component and all functional networks. 
Figure \ref{fig:allhist_comp_0}(a-c) shows how the number of (a) negative, (b) positive
and (c) all interactions is distributed. All three plots reveal a significant statistical difference
between networks in the wildtype component and the set of all functional networks. Networks in the wildtype component are sparse compared with the average functional network. 
 
Geometric information of the neutral graph is provided in Figure \ref{fig:allhist_comp_0}(d) 
in terms of the Hamming distance of functional networks from the wildtype. Functional networks
in the wildtype component are closer to the wildtype than the average functional
network is. Still the most remote networks in the wildtype component are found at distance 
$77$ from the wildtype. Despite its moderate size, the wildtype component pervades
a large part of the network space.

Shifting attention from the structural to the dynamical properties of the functional networks,
let us analyze the resilience of the dynamics against perturbations. As a measure of resilience
we use the $G_1$ basin size \cite{Li2004}, i.e.\ the number of states from which
the dynamics eventually reaches the fixed point $G_1$. Clearly, the basin contains at least 
the 13 states in the cell cycle sequence. As shown by the distributions in Figure
\ref{fig:allhist_comp_0}(e), actual $G_1$ basin sizes in functional networks contain many
more states. Compared with all functional networks, basin sizes of networks in the wildtype component
concentrate at higher values. The most frequently observed basin size is 2047 for networks in the 
wildtype component, cf.\ the inset of Figure \ref{fig:allhist_comp_0}(e). 
However, we have not found a functional network where the G1 basin contained all 2048 states.
\begin{figure*}
\centerline{\includegraphics[width=0.80\textwidth]{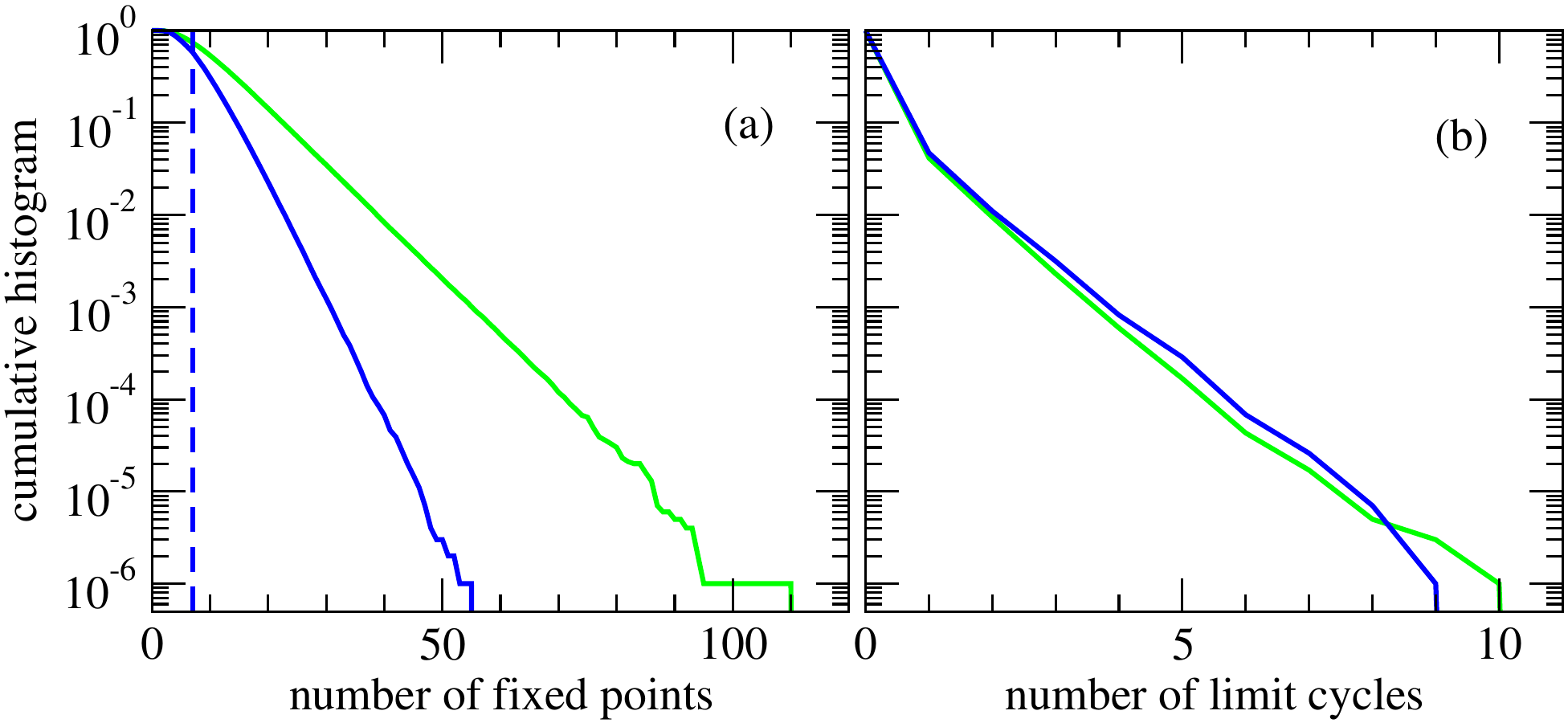}}
\caption{\label{fig:cumul_hist_attractors_cerev}       
The number of attractors of all functional networks (green curves) and the functional networks in
the neutral graph component containing the wildtype (blue curves). (a) Distribution of the number of fixed points.
(b) Distribution of the number of limit cycles (attractors of length at least 2).
The wildtype itself has 7 fixed points (vertical dashed line) and no limit cycles.
}
\end{figure*}
Moreover, the distributions in the number of fixed points of functional networks show a striking
difference between the wildtype component and the whole neutral graph. Figure
\ref{fig:cumul_hist_attractors_cerev}(a) displays geometric distributions in both cases.
However, networks in the wildtype component have a significantly narrower distribution of
fixed points. Interestingly, dynamic attractors (limit cycles) with more than one state
show practically the same statistics in the wildtype component as in the whole neutral
graph, cf. Figure \ref{fig:cumul_hist_attractors_cerev}(b).

\section{Computational aspects} \label{sec:computational}

As noted by Lau et al.\ \cite{Lau2007}, the set of network matrices performing a
given state sequence has a simple combinatorial structure. One can check
independently for each node $i$ if it takes the required state at each time step
$t$. The states taken by node $i$ only depend on the $i$-th row and not on the
whole matrix. Thus, a functional network can be constructed by independently
combined functional row vectors into a matrix. The set of  functional row
vectors for each node $i$ is obtained by testing each of the $3^{11} \approx 2
\times 10^5$ possible vectors over $\{-1,0,1\}$. For observables, such as the
number of interactions and Hamming distances that fall into sums over row
vectors of the matrix, exact distributions are obtained by combining the
distributions for all rows. In fact, each row of the matrix has its own neutral
graph. The Cartesian product \cite{Imrich2000} of these is the neutral graph of
the whole system. Sampling is used to obtain statistics of observables that are
not a function of single rows, such as the number of attractors and basin
sizes.  

\section{Discussion \& Outlook} \label{sec:discussion}
We have analyzed the {\em neutral graph} (also called neutral network) of discrete regulatory
networks reproducing the cell cycle sequence of budding yeast \cite{Li2004}. The neutral graph falls
into many connected components. Networks in different components of the neutral
graph are not accessible to each other through a sequence of mutations that retains cell cycle
functionality. Our finding contrasts with the connected neutral graphs in the work by Ciliberti
et al.\ in a similar type of discrete regulatory networks \cite{Ciliberti2007a,Ciliberti2007b}. 
There, function is defined as the eventual arrival at a predefined fixed point
from a given initial condition. In the present study, the exact sequence of states leading to the
fixed point is part of the required phenotype. We hypothesize that the fragmentation of the neutral
graph is caused by increasing functional constraints.

Further analysis has revealed that functional networks accessible from the empirical wildtype
are structurally and dynamically distinct from other functional networks. 
Networks in the wildtype component are more sparsely wired and their dynamics is more resilient 
to perturbations, as compared to the average of all functional networks. 

Thus, networks in the wildtype component have properties similar to the wildtype itself. This is remarkable since most networks in the wildtpye component are distant from the wildtype, having only a few interactions in common.

Future investigations could establish conditions for the connectedness of the neutral graph.
To what extent is the fragmentation of the neutral graph caused by the strong discretization of
interaction strengths? Allowing finer adaptations would lead to less fragmented neutral graphs. 
In the extreme (though chemically unrealistic) limit of continuously evolving interaction strengths, 
the set of all functional network matrices is convex and thus connected. 

\begin{acknowledgements}
We thank Anke Busch, Nadine Menzel, and Markus Riester for valuable comments on the draft.
This work has been funded by the VolkswagenStiftung.
\end{acknowledgements}

\bibliographystyle{spphys}       
\bibliography{Boldhaus}   

\end{document}